\documentclass[conference]{IEEEtran}
\IEEEoverridecommandlockouts
\usepackage{cite}
\usepackage{amsmath,amssymb,amsfonts}
\usepackage{algorithmic}
\usepackage{graphicx}
\usepackage{textcomp}
\makeatletter

\newcommand{\Rmnum}[1]{\expandafter\@slowromancap\romannumeral #1@}
\makeatother
\usepackage{multirow, makecell}
\def\BigRoman{\uppercase\expandafter{\romannumeral\number\count 255 }}
\def\Romannumeral{\afterassignment\BigRoman\count255=}
\usepackage{tikz}
\def\BibTeX{{\rm B\kern-.05em{\sc i\kern-.025em b}\kern-.08em
    T\kern-.1667em\lower.7ex\hbox{E}\kern-.125emX}}

\begin{document}

\title{Towards Neurohaptics: Brain-Computer Interfaces for Decoding Intuitive Sense of Touch
\footnote{\thanks{20xx IEEE. Personal use of this material is permitted. Permission
from IEEE must be obtained for all other uses, in any current or future media, including reprinting/republishing this material for advertising or promotional purposes, creating new collective works, for resale or redistribution to servers or lists, or reuse of any copyrighted component of this work in other works.

This work was partly supported by Institute of Information \& Communications Technology Planning \& Evaluation (IITP) grant funded by the Korea government (MSIT) (No. 2017-0-00432, Development of Non-Invasive Integrated BCI SW Platform to Control Home Appliances and External Devices by User’s Thought via AR/VR Interface; No. 2017-0-00451, Development of BCI based Brain and Cognitive Computing Technology for Recognizing User’s Intentions using Deep Learning; No. 2019-0-00079, Artificial Intelligence Graduate School Program (Korea University)).}
}
}

\author{\IEEEauthorblockN{Jeong-Hyun Cho}
\IEEEauthorblockA{\textit{Dept. Brain and Cognitive Engineering}\\
\textit{Korea University} \\
Seoul, Republic of Korea \\
jh\_cho@korea.ac.kr}\\

\IEEEauthorblockN{Myoung-Ki Kim}
\IEEEauthorblockA{\textit{Dept. Artificial Intelligence}\\
\textit{Korea University} \\
Seoul, Republic of Korea \\
\textit{LG Display} \\
kim\_mk@korea.ac.kr}\\

\and

\IEEEauthorblockN{Ji-Hoon Jeong}
\IEEEauthorblockA{\textit{Dept. Brain and Cognitive Engineering}\\
\textit{Korea University} \\
Seoul, Republic of Korea \\
jh\_jeong@korea.ac.kr}\\

\IEEEauthorblockN{Seong-Whan Lee}
\IEEEauthorblockA{\textit{Dept. Artificial Intelligence}\\
\textit{Korea University} \\
Seoul, Republic of Korea \\
sw.lee@korea.ac.kr}
}

%\author{\IEEEauthorblockN{}
%\IEEEauthorblockA{{$^1$Department of Brain and Cognitive Engineering, Korea University, Seoul, Republic of Korea} \\
%{$^2$Department of Artificial Intelligence, Korea University, Seoul, Republic of Korea}\\
%}
%}

\maketitle

\begin{abstract}
Noninvasive brain--computer interface (BCI) is widely used to recognize users' intentions. Especially, BCI related to tactile and sensation decoding could provide various effects on many industrial fields such as manufacturing advanced touch displays, controlling robotic devices, and more immersive virtual reality or augmented reality. In this paper, we introduce haptic and sensory perception-based BCI systems called neurohaptics. It is a preliminary study for a variety of scenarios using actual touch and touch imagery paradigms. We designed a novel experimental environment and a device that could acquire brain signals under touching designated materials to generate natural touch and texture sensations. Through the experiment, we collected the electroencephalogram (EEG) signals with respect to four different texture objects. Seven subjects were recruited for the experiment and evaluated classification performances using machine learning and deep learning approaches. Hence, we could confirm the feasibility of decoding actual touch and touch imagery on EEG signals to develop practical neurohaptics.
\end{abstract}

\begin{small}
\textbf{\textit{Keywords-brain--computer interface; electroencephalogram; tactile information; haptic sensation analysis; touch imagery}}\\
\end{small}

\section{Introduction}
Brain--computer interface (BCI)s are systems that could decode brain signals to understand the intention and status of people. BCIs are also used to analyze various tactile sensations related to haptics. Numerous studies have attempted to understand electroencephalogram (EEG) signals because the signals contain significant information about the cognition of people \cite{C3,MRCP,B1,ECoG2}. For decades, EEG--based BCIs focused and investigated into several paradigms for signal acquisition such as movement-related cortical potential \cite{jeong2020decoding,MRCP}, event-related potential (ERP) \cite{EEG,A1}, and motor imagery \cite{A2,C2,kam}. As applications of the BCIs, robotic arm controls \cite{A2}, speller systems \cite{speller,won2017motion}, brain--controlled wheelchairs \cite{wheelchair}, and the neurohaptics \cite{osborn2018prosthesis, ganzer2020restoring, tayeb2020decoding} were commonly used for communication between human and machines. However, the neurohaptics are relatively few compared to conventional paradigms.

%%%%%%%%%%%%%%%%%%%%%%%%%%%%%%%%%%%%%%%%%%%%%%%%%%%%%%%%%%%%%%%%%%%%%%
\begin{figure*}[t]
\centerline{\includegraphics[width=\textwidth]{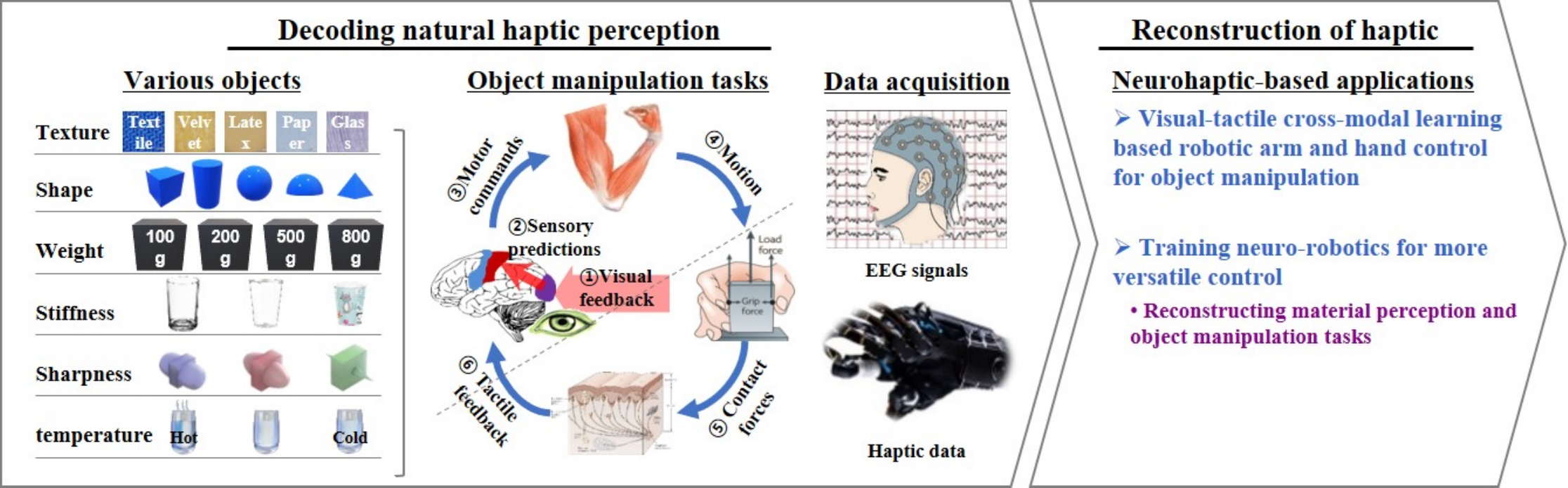}}
\caption{Overview of decoding natural haptic perception and reconstructing the haptic feedback to develop haptic-related BCI systems.}
\end{figure*}
%%%%%%%%%%%%%%%%%%%%%%%%%%%%%%%%%%%%%%%%%%%%%%%%%%%%%%%%%%%%%%%%%%%%%

Recently, a BCI-based haptic sensation is one of the most interesting topics. The haptic sensation is an electrical sense of objects and surfaces felt through the nerves in the skin of the fingers. Before haptic sensation studies became active in the non-invasive BCI research field, invasive BCI-based haptic studies had begun first. Osborn \emph{et al}. \cite{osborn2018prosthesis} developed a haptic feedback system that operates on a prosthetic basis. Based on the tactile information obtained through the electronic skin installed on the robot's prosthetic arm and the sensory recognition information of the subjects obtained through the BCI, the sensory information felt when holding a sharp object was made available to the brain by using a transcutaneous nerve stimulation that could give electrical stimulation to the nerves. Ganzer \emph{et al}.\cite{ganzer2020restoring} also implemented a bi-directional communication system of the brain and haptic feedback. Based on brain signals from invasive BCI, they developed a system that could give patients adequate electrical feedback through a functional electrical stimulation (FES) attached to the arm. When the patient with reduced finger sensation holds the object with a hand, the system amplified the haptic perception to assist the patient makes a better and stronger grasp through FES. Tayeb \emph{et al}. \cite{tayeb2020decoding} proposed a non-invasive BCI system that could implement the same system as the studies described above. They developed a prosthetic arm that can restore the sense of touch and pain. Brain responses are analyzed and decoded to understand the tactile sensory perception, including pain, and identify activated brain regions. Neural activity can be used to design a prosthesis that mimics natural pain withdrawal behavior in humans. The overview in Fig. 1 illustrates what tactile information is being targeted to analyze in recent studies and what purpose and process are using EEG and haptic information.

In this study, we measured EEG signals of 4-class of actual touch and touch imagery (`Fabric', `Glass', `Paper', and `Fur'). The classes used in the experimental paradigm consist of the most basic haptic perception for analysis of touch sensation. To the best of our knowledge, this is the first attempt that demonstrates the feasibility of classifying the high-level haptic perception which consists of 4-class to develop a high-performance classification model based on a deep learning approach. Second, we achieved robust classification performance in the 4-class touch perception compared with the chance-level accuracy (0.25).

\section {Materials and Methods}
\subsection{Participants}
Seven healthy subjects, who were naive BCI users, have recruited in the experiment (aged 25-38, all right--handed). Before the experiment, each subject was informed of the experimental protocols, paradigms, and purpose. After they had understood, all of them provided their given written consent according to the Declaration of Helsinki. The experimental protocols and environments were reviewed and approved by the Institutional Review Board at Korea University [KUIRB-2020-0013-01].

%%%%%%%%%%%%%%%%%%%%%%%%%%%%%%%%%%%%%%%%%%%%%%%%%%%%%%%%%%%%%%%%%%%%%%
\begin{figure}[ht!]
\centerline{\includegraphics[width=\columnwidth]{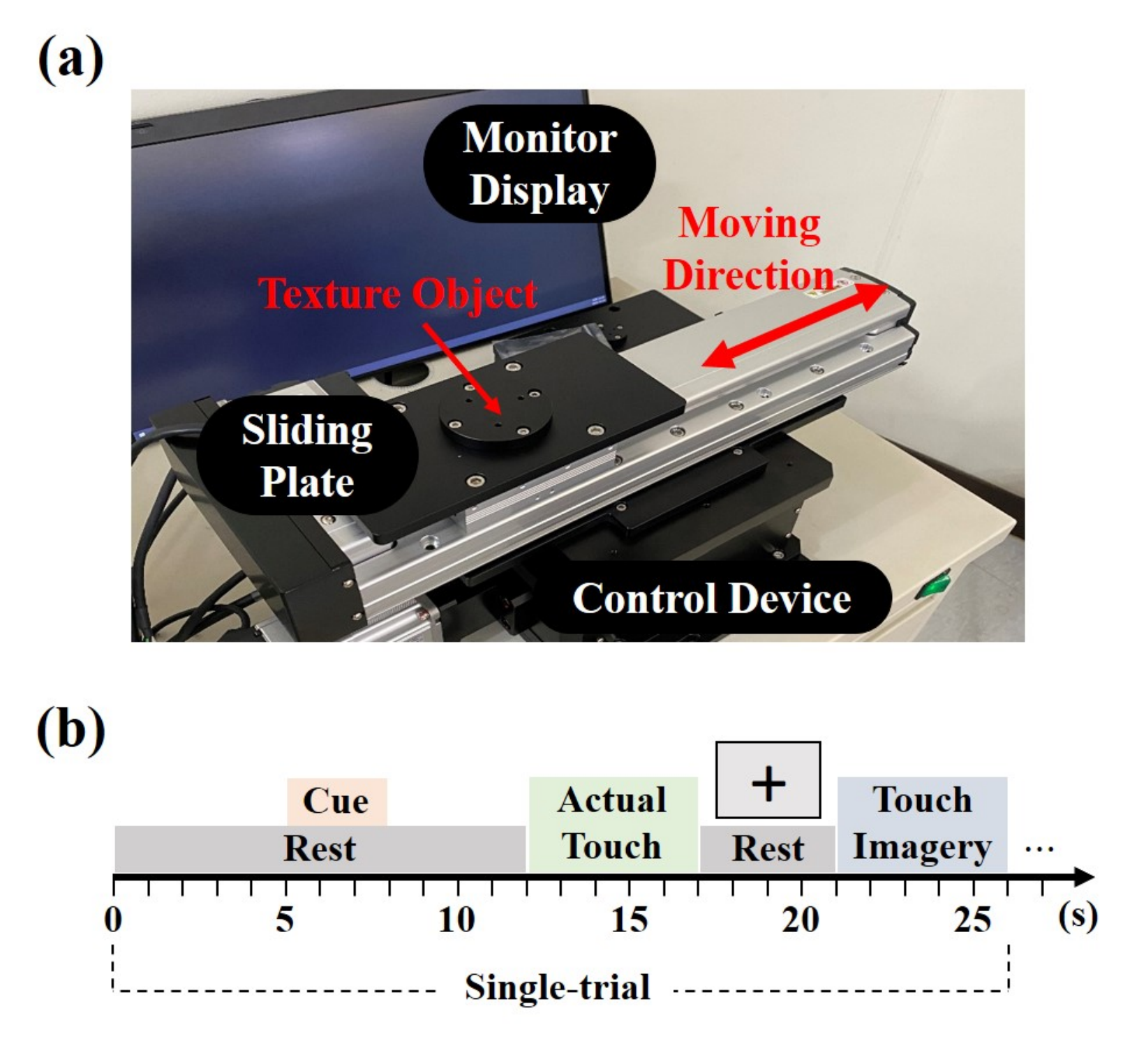}}
\caption{Experimental setup and protocol to record touch sensation. A self-made instrument for the purpose of obtaining EEG signals under consistent tactile sensation (a) and experimental protocol for recording EEG signals related to the actual touch and the touch imagery (b).}
\end{figure}
%%%%%%%%%%%%%%%%%%%%%%%%%%%%%%%%%%%%%%%%%%%%%%%%%%%%%%%%%%%%%%%%%%%%%

\subsection{Experimental Setup}
First, a monitor display for touch instruction was put at a distance of 90cm from the subjects. At the same time, we devised a control device that can give the subject a consistent and uniformed tactile sensation with his hands fixed as depicted in Fig. 2 (a). The subjects were asked to perform actual touch and touch imagery according to the four different tactile objects. Each trial was composed of four phases such as rest, cue/preparation, actual touch, and imagery, as shown in Fig. 2 (b). In the rest phase, the subject took a comfortable rest with restraining eye and body movement. Between the rest phase, the monitor displays one of the texture images (picture of the four texture objects) as a touch cue, and then subjects prepare the touch imagery task according to the cue. The subjects stared the fixation point during 3 s to avoid an afterimage effect. During the 5 s, The subjects conducted a touch imagery task. We asked subjects to perform 200 trials in total (i.e., 50 trials $\times$ 4 classes).

\subsection{EEG Signal Acquisition}
We acquired the EEG signals via BrainVision Recorder (BrainProducts GmbH, Germany). EEG signals were acquired using 64 Ag/AgCl electrodes following 10/20 international systems. The ground and reference channels were FCz and FPz positions, respectively. The sampling rate was 1,000 Hz, and a notch filter was applied to the acquired signals as 60 Hz. All electrode impedances were kept below 10 k$\Omega$ during the experiment.

\subsection{Control Device for Tactile Sensation}
We designed a device for the experiment that could provide the subjects with a consistent and uniform texture feeling. As shown in Fig. 2 (a), the control device consists of a rail and a plate that travels horizontally above it. The subject's finger was able to feel the texture of the objects without moving itself according to the experimental protocol. The actual touch was conducted by touching four different tactile objects which are 'Fabric', `Glass', `Paper', and `Fur.' The fabric cue indicated touching a smooth surface for the haptic sensation. In the touch imagery phase, the subjects imagined the fabric texture that they experienced in the actual touch phase. The glass cue shows that the hard and slippy surface for the texture sensations. The paper cue presents the surface as smooth but relatively not slippy compares to the glass texture. Finally, the fur cue indicates the texture which is smoother than the fabric class, and this surface texture has a characteristic of high friction because of the structure of the raw material.

\subsection{Data Analysis}
For data analysis, we compared the algorithm of common spatial patterns (CSP)--linear discriminant analysis (LDA), a popular method for motor and haptic-related EEG decoding, and the architecture of EEGNet, which shows remarkable performance in classification tasks. We segmented the data into 5 s epoched data for each trial so we could prepare 5 s of actual touch and 5 s of touch imagery from a single trial. Then, the CSP algorithm \cite{FBCSP} was applied to extract dominant spatial features for training. A transformation matrix from CSP consisted of the logarithmic variances of the first three and the last three columns were used as a feature. The LDA \cite{channel} was used for a classification method which classified four different class using one-versus-rest strategy. EEGNet which is a convolutional neural network (CNN) architecture proposed by Lawhern \emph{et al} used the preprocessed EEG data as input \cite{lawhern2018eegnet}. No conventional feature extraction, such as CSP, was performed and only significant features were expected to be extracted with the operation of the convolution layer. For a fair evaluation of classification performance, 5-by-5-fold cross-validation was used.

%%%%%%%%%%%%%%%%%%%%%%%%%%%%%%%%%%%%%%%%%%%%%%%%%%%%%%%%%%%%%%%%%%%%%%%%%%%%%%%%%%%%%%%
\renewcommand{\arraystretch}{1.1}
\begin{table}[t!]
\tiny
\centering
\caption{4-Class Classification Accuracy in Actual Touch and Touch Imagery}
\label{table1}
\resizebox{\columnwidth}{!}{%
\begin{tabular}{ccccc}
\hline
\multicolumn{1}{l}{\multirow{3}{*}{\textbf{}}} &
  \multicolumn{4}{c}{\textbf{\begin{tabular}[c]{@{}c@{}}4-Class classification accuracy\\ (5-by-5-fold cross-validation)\end{tabular}}} \\ \cline{2-5} 
\multicolumn{1}{l}{} &
  \multicolumn{2}{c}{\textbf{Actual touch}} &
  \multicolumn{2}{c}{\textbf{Touch imagery}} \\
\multicolumn{1}{l}{} &
  \textbf{CSP-LDA} &
  \textbf{EEGNet} &
  \textbf{CSP-LDA} &
  \textbf{EEGNet} \\ \hline
\textbf{Sub 1} &
  \begin{tabular}[c]{@{}c@{}}0.5194\\ ($\pm$0.0294)\end{tabular} &
  \begin{tabular}[c]{@{}c@{}}0.6475\\ ($\pm$0.1021)\end{tabular} &
  \begin{tabular}[c]{@{}c@{}}0.4240\\ ($\pm$0.0350)\end{tabular} &
  \begin{tabular}[c]{@{}c@{}}0.5506\\ ($\pm$0.0433)\end{tabular} \\
\textbf{Sub 2} &
  \begin{tabular}[c]{@{}c@{}}0.5465\\ ($\pm$0.0503)\end{tabular} &
  \begin{tabular}[c]{@{}c@{}}0.6229\\ ($\pm$0.0935)\end{tabular} &
  \begin{tabular}[c]{@{}c@{}}0.4793\\ ($\pm$0.1305)\end{tabular} &
  \begin{tabular}[c]{@{}c@{}}0.5241\\ ($\pm$0.0637)\end{tabular} \\
\textbf{Sub 3} &
  \begin{tabular}[c]{@{}c@{}}0.4599\\ ($\pm$0.1113)\end{tabular} &
  \begin{tabular}[c]{@{}c@{}}0.6202\\ ($\pm$0.1208)\end{tabular} &
  \begin{tabular}[c]{@{}c@{}}0.3588\\ ($\pm$0.0751)\end{tabular} &
  \begin{tabular}[c]{@{}c@{}}0.4122\\ ($\pm$0.0116)\end{tabular} \\
\textbf{Sub 4} &
  \begin{tabular}[c]{@{}c@{}}0.6631\\ ($\pm$0.0891)\end{tabular} &
  \begin{tabular}[c]{@{}c@{}}0.7191\\ ($\pm$0.0813)\end{tabular} &
  \begin{tabular}[c]{@{}c@{}}0.5105\\ ($\pm$0.0385)\end{tabular} &
  \begin{tabular}[c]{@{}c@{}}0.5174\\ ($\pm$0.0181)\end{tabular} \\
\textbf{Sub 5} &
  \begin{tabular}[c]{@{}c@{}}0.6608\\ ($\pm$0.0699)\end{tabular} &
  \begin{tabular}[c]{@{}c@{}}0.6615\\ ($\pm$0.0732)\end{tabular} &
  \begin{tabular}[c]{@{}c@{}}0.3092\\ ($\pm$0.0407)\end{tabular} &
  \begin{tabular}[c]{@{}c@{}}0.2935\\ ($\pm$0.0974)\end{tabular} \\
\textbf{Sub 6} &
  \begin{tabular}[c]{@{}c@{}}0.4674\\ ($\pm$0.0871)\end{tabular} &
  \begin{tabular}[c]{@{}c@{}}0.6461\\ ($\pm$0.0942)\end{tabular} &
  \begin{tabular}[c]{@{}c@{}}0.3811\\ ($\pm$0.0547)\end{tabular} &
  \begin{tabular}[c]{@{}c@{}}0.5603\\ ($\pm$0.0612)\end{tabular} \\
\textbf{Sub 7} &
  \begin{tabular}[c]{@{}c@{}}0.4155\\ ($\pm$0.1064)\end{tabular} &
  \begin{tabular}[c]{@{}c@{}}0.6366\\ ($\pm$0.0371)\end{tabular} &
  \begin{tabular}[c]{@{}c@{}}0.5347\\ ($\pm$0.1009)\end{tabular} &
  \begin{tabular}[c]{@{}c@{}}0.4112\\ ($\pm$0.1179)\end{tabular} \\
\textbf{\begin{tabular}[c]{@{}c@{}}Average\\ ($\pm$Std.)\end{tabular}} &
  \textbf{\begin{tabular}[c]{@{}c@{}}0.5332\\ ($\pm$0.0975)\end{tabular}} &
  \textbf{\begin{tabular}[c]{@{}c@{}}0.6506\\ ($\pm$0.0335)\end{tabular}} &
  \textbf{\begin{tabular}[c]{@{}c@{}}0.4282\\ ($\pm$0.0836)\end{tabular}} &
  \textbf{\begin{tabular}[c]{@{}c@{}}0.4670\\ ($\pm$0.0981)\end{tabular}} \\ \hline
\end{tabular}
}
\end{table}
%%%%%%%%%%%%%%%%%%%%%%%%%%%%%%%%%%%%%%%%%%%%%%%%%%%%%%%%%%%%%%%%%%%%%%%%%%%%%%%%%%%%%%%
%%%%%%%%%%%%%%%%%%%%%%%%%%%%%%%%%%%%%%%%%%%%%%%%%%%%%%%%%%%%%%
\begin{figure}[t]
\centerline{\includegraphics[width=\columnwidth]{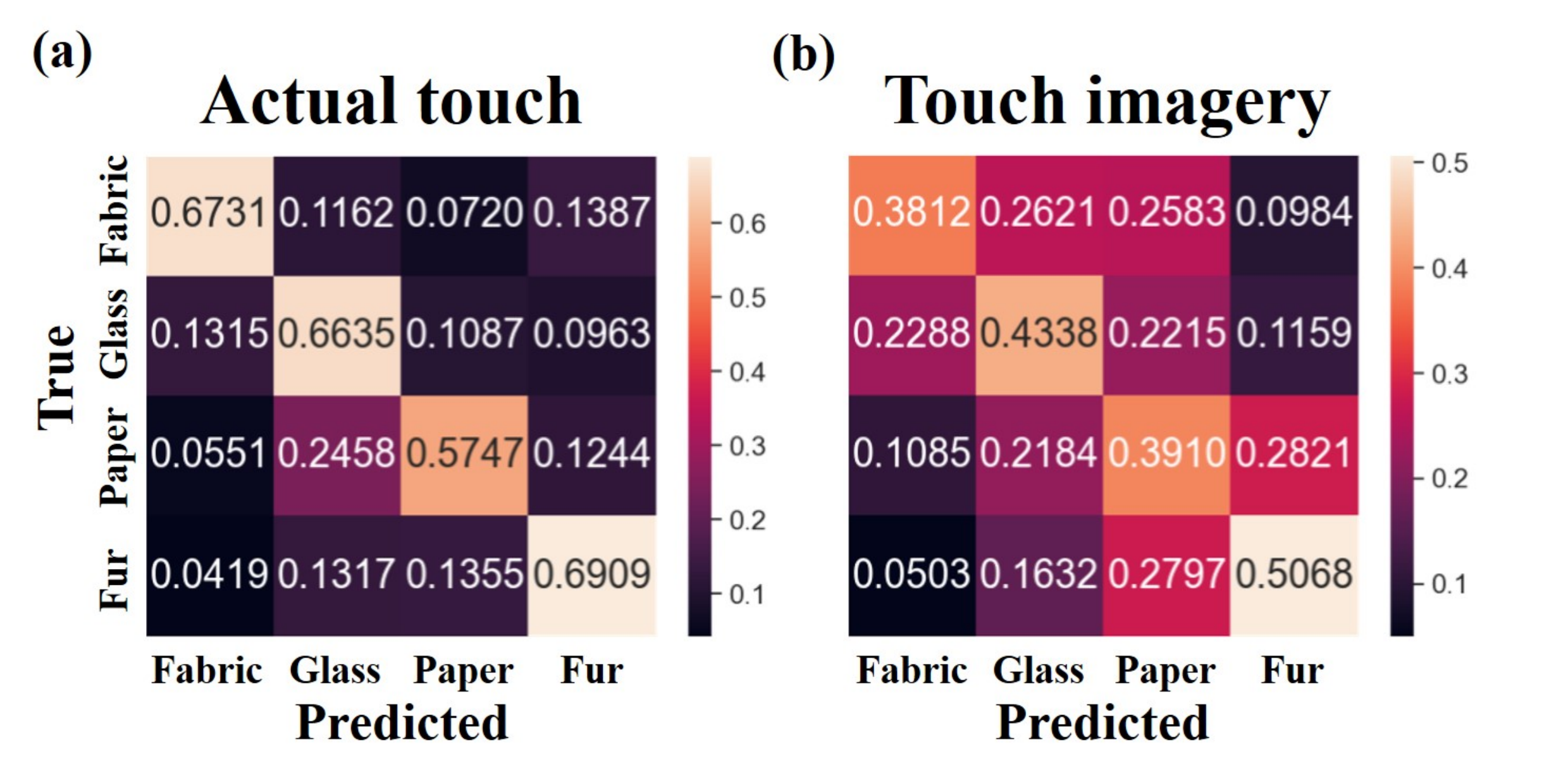}}
\caption{Confusion matrices of predicted and true labels to the corresponding classes. Result on actual touch (a) and result on touch imagery (b).}
\end{figure}
%%%%%%%%%%%%%%%%%%%%%%%%%%%%%%%%%%%%%%%%%%%%%%%%%%%%%%%%%%%%%%
%%%%%%%%%%%%%%%%%%%%%%%%%%%%%%%%%%%%%%%%%%%%%%%%%%%%%%%%%%%%%%
\begin{figure*}[t]
\centerline{\includegraphics[width=\textwidth]{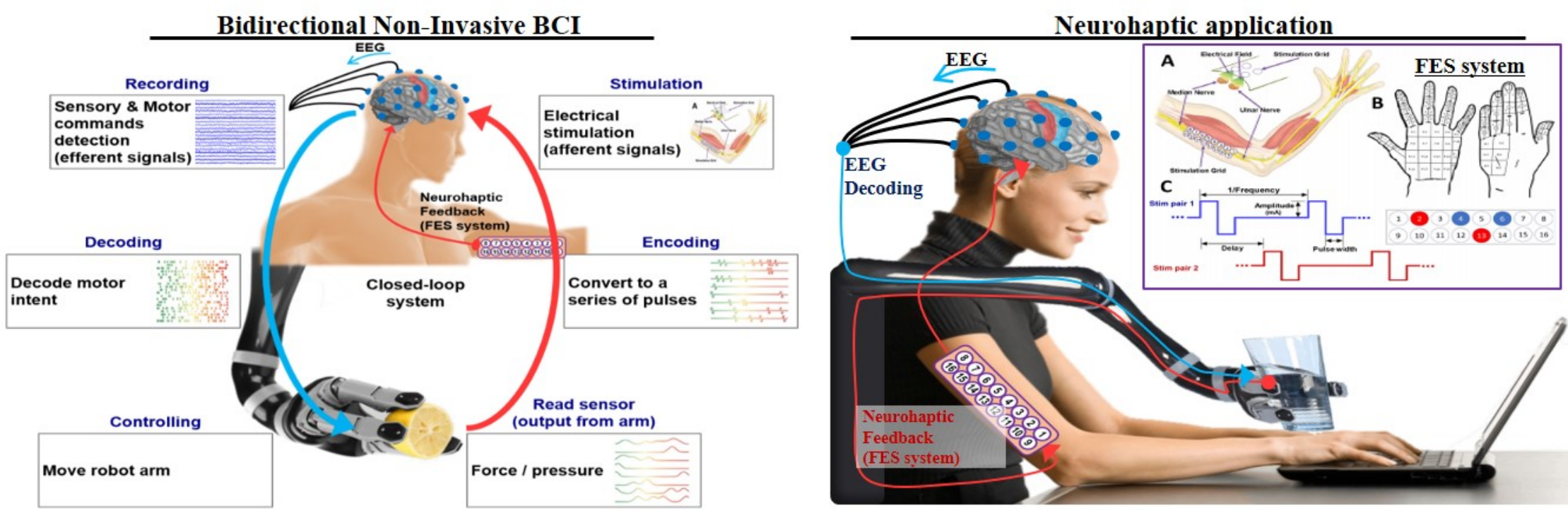}}
\caption{Illustration of describing bidirectional non-invasive BCI based neurohaptic system and its application.}
\end{figure*}
%%%%%%%%%%%%%%%%%%%%%%%%%%%%%%%%%%%%%%%%%%%%%%%%%%%%%%%%%%%%%%

\section {Results and Discussion}
As shown in Table I, we conducted an experiment of actual touch and touch imagery on seven subjects for 4-class classification. As a result, classification using EEGNet performed better than the CSP-LDA method, especially after analyzing actual touch, it was found that using EEGNet improved average performance by about 0.1174. The difference in classification performance between actual touch and touch imagery was lower than expected. For example, the difference between the average accuracy of the actual touch and touch imagery classification is not significant (0.1050--0.1836).

In Fig. 3, the confusion matrices for the average classification performance of EEGNet were presented. Fig. 3 (a) is for actual touch and Fig. 3 (b) is for touch imagery. We could confirm the decoding model could perform classification clearly on actual touch and its performance is relatively less effective on touch imagery. At the same time, we can see that classifying the texture of glass and paper are more confusing than other classes. On the other hand, the texture of fabric and fur are relatively distinguishable. 

\section{Conclusion and Future works}
In this paper, we designed an experimental environment for acquiring EEG data with respect to touch imagery. Through the experiment system, the subjects could perform the touch imagery for the haptic analysis of various tactile information. We have implemented the four classes representing natural texture. The classification of touch imagery could contribute to developing neurohaptic systems for industry, VR/AR application, and artificial intelligence. Fig. 4 shows an overview of the BCI-based bi-directional haptic system that we will challenge in the future. The purpose is to provide real-time haptic feedback, along with analysis of motor intention to assist object cognition and application controllability of users for the tasks needed in everyday life. 

As a result, the EEG classification performance needs to be higher. We will continue to test and adopt advanced deep learning approaches to improve our haptic-related BCI system robust in real-world environments. It would greatly improve the interaction effect between the user and BCIs.

\bibliographystyle{IEEEbib}
\bibliography{refs}

\end{document}